# Reentrant Superconductivity and Superconducting Critical Temperature Oscillations in F/S/F trilayers of Cu$_{41}$Ni$_{59}$/Nb/Cu$_{41}$Ni$_{59}$ Grown on Cobalt Oxide

V.I. Zdravkov[1,2], J. Kehrle[1], D. Lenk[1], G. Obermeier[1], A. Ullrich[1], C. Müller[1], H.-A. Krug von Nidda[1], R. Morari[2], A.S. Sidorenko[2], L.R. Tagirov[1,3], S. Horn[1], R. Tidecks[1]

[1]*Institut für Physik, Universität Augsburg, D-86159 Augsburg, Germany*

[2]*D. Ghitsu Institute of Electronic Engineering and Nanotechnologies ASM, MD2028 Kishinev, Moldova*

[3]*Solid State Physics Department, Kazan Federal University, 420008 Kazan, Russia*

**Abstract**

Ferromagnet/Superconductor/Ferromagnet (F/S/F) trilayers constitute the core of a superconducting spin valve. The switching effect of the spin valve is based on interference phenomena occurring due to the proximity effect at the S/F interfaces. A remarkable effect is only expected if the core structure exhibits strong critical temperature oscillations, or most favorable, reentrant superconductivity, when the thickness of the ferromagnetic layer is increased. The core structure has to be grown on an antiferromagnetic oxide layer (or such layer to be placed on top) to pin by exchange bias the magnetization-orientation of one of the ferromagnetic layers. In the present paper we demonstrate that this is possible, keeping the superconducting behavior of the core structure undisturbed.

**I. Introduction**

In a superconducting spin valve [1,2], the critical temperature of a superconductor (S) sandwiched by two ferromagnets (F) depends on the relative orientation of the magnetization of the F-layers. To realize such device, fine tuning of material parameters and fabrication technology of F/S/F trilayers, representing the core structure, is necessary. Recently, we succeeded in the realization of a Cu$_{41}$Ni$_{59}$/Nb/Cu$_{41}$Ni$_{59}$ trilayer [3] exhibiting the unusual non-monotonic behavior of the transition temperature $T_c$, required for the functioning of a spin valve device. Due to the establishing of a Fulde-Ferrell [4] Larkin-Ovchinnikov [5] (FFLO) like state, interference effects of the superconducting pairing wave function [6] lead to an oscillatory behavior of $T_c$ and reentrant superconductivity, *i.e.* an extinction and recovery of the superconducting state when increasing the F-layers thickness. These phenomena were previously extensively studied in superconductor-ferromagnet bilayers of the same materials [7 – 9]. As a next technological step one has to demonstrate that the core structure with the required properties can be fabricated with one of the magnetic layers in contact with an antiferromagnet. The reason is that to enable the adjustment of parallel and antiparallel magnetizations alignment of the F-layers in F/S/F trilayers by an external magnetic field, one needs to exchange bias one F-layer with an antiferromagnet [10]. Therefore, in this work we have grown the F/S/F core structure on top of an antiferromagnetic (AF) material (or placed an AF layer on top of the trilayer). In the following we report on the fabrication and the superconducting properties of such stacks on cobalt oxide layer, keeping deep critical temperature oscillations or reentrant superconductivity, which is necessary to get a large size of the spin-valve effect [8].

## II. Sample Preparation and Characterization
### A. Thin Film Deposition

The samples were deposited by magnetron sputtering with a Leybold Z400 system, usually on Si (111) substrates. Argon gas (purity 99.999%, pressure $8 \times 10^{-3}$ mbar) was used for deposition of non-oxide layers whereas to form cobalt oxide layer we used the mixture of argon and oxygen with flow rate 5:1. Three targets can be mounted in the system simultaneously, therefore, we interrupted the fabrication of heterostructures after deposition of the Co oxide layer and replaced the Co target by a $Cu_{40}Ni_{60}$ one. About 30 samples have been prepared in one run by applying a wedge technique [7, 8] and cutting the 80 mm long and 7-8 mm wide specimen (typical dimensions) in stripes of equal width, perpendicular to the thickness gradient of the wedge. In Fig. 1 sketches of spin valve structures are shown.

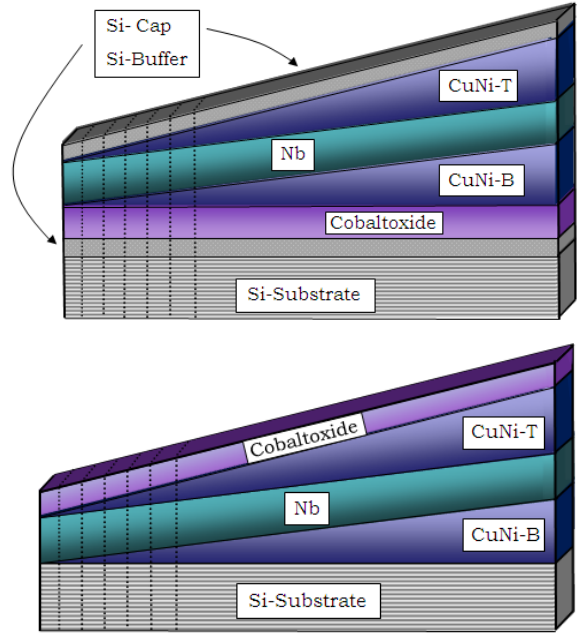

**Fig. 1**: Schematic diagram of an AF-FSF (top panel) and FSF-AF (bottom panel) spin valve structure, respectively. Here, CuNi-B and CuNi-T denote the bottom and top ferromagnetic layer. The superconducting layer is made of niobium. The antiferromagnetic (AF) cobalt oxide serves to exchange bias the CuNi-B layer in the AF-FSF structure and the CuNi-T layer in the FSF-AF structure, respectively. For the FSF-AF system there is no Si cap and buffer layer.

First, the targets were pre-sputtered in order to remove any adsorbents from the surface. In the case of the AF-FSF structure (with the AF-layer at the bottom) a buffer layer of amorphous silicon was then deposited. To remove any contaminations from the surface of the cobalt-oxide layer after breaking of vacuum it was pre-sputtered, then, a wedge of ferromagnetic alloy (the resulting composition according to Rutherford Backscattering Spectrometry (RBS) is usually about 41 at.% Cu and 59 at.% Ni as studied for S/F and F/S bilayers [7 – 9]) was deposited followed by a niobium layer of constant thickness (applying a spray technique [7, 8], purity of the niobium target 99.99 %) and a $Cu_{41}Ni_{59}$ wedge. The functional layers were protected from ambient conditions by a silicon cap.

Sample series AF-FSF4 was prepared at room temperature. To improve the growth of the cobalt oxide layer, the substrate was heated to 300°C during the deposition of this layer for sample series AF-FSF5a. The other layers in these series were prepared at room temperature.

If the antiferromagnetic layer is placed on the top of the stack, neither a Si buffer nor a Si cap (to protect the system against ambient conditions) was deposited, to sputter all layers without breaking the vacuum to change the targets. Using this method, sample series FSF-AF2 was prepared (heating during the deposition of the top $Cu_{41}Ni_{59}$ layer and the Co oxide layer to 200°C, rest prepared at room temperature).

**B. Thickness Determination**

In order to determine the thickness of the individual layers, Rutherford Backscattering Spectrometry was used following Refs. [3, 7 – 9]. Thereby, α-particles with an energy of 3.5MeV are accelerated onto the specimen. The energy of the backscattered helium atoms is detected under an angle of 170° with respect to the incoming beam. The sample was tilted by 7° in order to avoid channeling effects of the helium atoms in the mono-crystalline Si-substrate. Since the energy loss of the helium atoms when scattered elastically with copper, nickel and cobalt is very similar, the signals overlap in the spectrum, as shown for sample #2 of the AF-FSF5a series in Fig. 2.

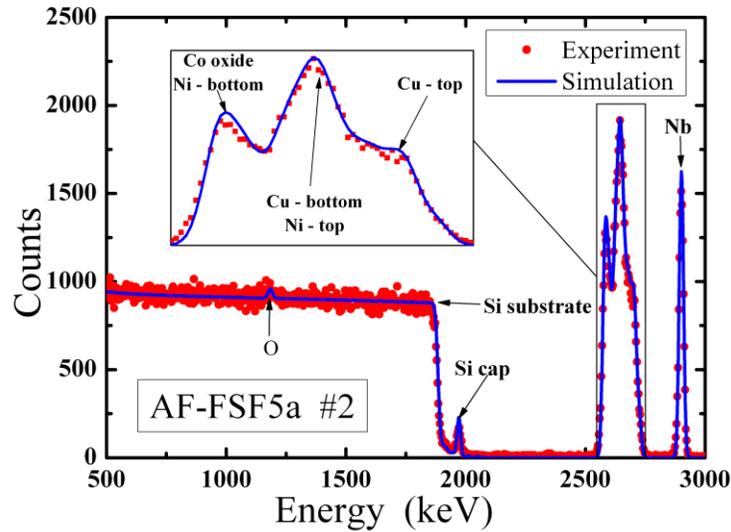

**Fig.2:** RBS measurement of sample AF-FSF5a #2. The cobalt oxide layer is deposited on the Si substrate (covered by a Si buffer layer) on which the F/S/F structure is placed. The functional layers are protected from ambient conditions by a Si cap. The cobalt in the cobalt-oxide layer and the two ferromagnetic layers give rise to one large peak.

Therefore, it is not possible to determine the thickness of the individual layers definitely, as it can be done in the case of S/F and F/S bilayers [8, 9] and F/S/F trilayers [3]. The thicknesses of the cobalt oxide layer and CuNi alloy layers were determined by Transmission Electron Microscope (TEM) investigations in addition. Since the RBS measurements of bilayers and trilayers show that uniformly fabricated layers indeed have a constant thickness over the whole series, it is reasonable to assume the thickness of the cobalt oxide layer, as determined by the TEM, to be present in the whole series as a first approximation. Moreover, the atomic composition of the ferromagnetic alloy was kept constant at about 41 at. % Cu and 59 at. % Ni. For the cobalt oxide layer it was found that the fitting can be better done with an atomic concentration of Co with 43 at. % and O with 57 at. % than with both equal to 50 at. %. Since 43/57 is about 3/4, it is concluded that the cobalt oxide in the sample seems to be rather $Co_3O_4$ than CoO or at least a mixture of both oxides with considerable amount of $Co_3O_4$. High resolution TEM (HRTEM) analysis discussed in Section II.C also gives evidence for $Co_3O_4$ as well as for CoO.

We began the fitting from the topmost layer, the silicon cap layer. Afterwards, the thickness of the top ferromagnetic layer was determined by fitting the right hand side of the niobium layer peak, which is shifted to lower energies by inelastic scattering of the helium ions in the top $Cu_{41}Ni_{59}$ layer. The thickness of the niobium film was determined by adjusting the full width half maximum of its peak. The thickness of the bottom ferromagnetic layer was obtained by matching the fitting curve to the peak arising from the copper atoms in this layer, as well as the left side of the peak arising from the nickel atoms in this layer.

The layer thicknesses, derived from the areal densities of the elements, and this fitting procedure for sample series AF-FSF4 are summarized in Fig. 3, top panel. Two ferromagnetic layers, with about the same wedge thickness profile, enclose a niobium layer with an average thickness of 13.4 nm. The silicon cap layer and cobalt oxide layer have an average thickness of 11.5 nm and 7.7 nm, respectively.

Using the same procedure to fit the RBS spectra of sample series AF-FSF5 a, the respective thicknesses were obtained (Fig. 3, middle panel). The two ferromagnetic layers in this case have practically the same wedge thickness profile. The niobium layer in between has an average thickness of 11.1 nm. The thicknesses of the silicon cap and cobalt oxide layer are 11.3 nm and 8.4 nm, respectively.

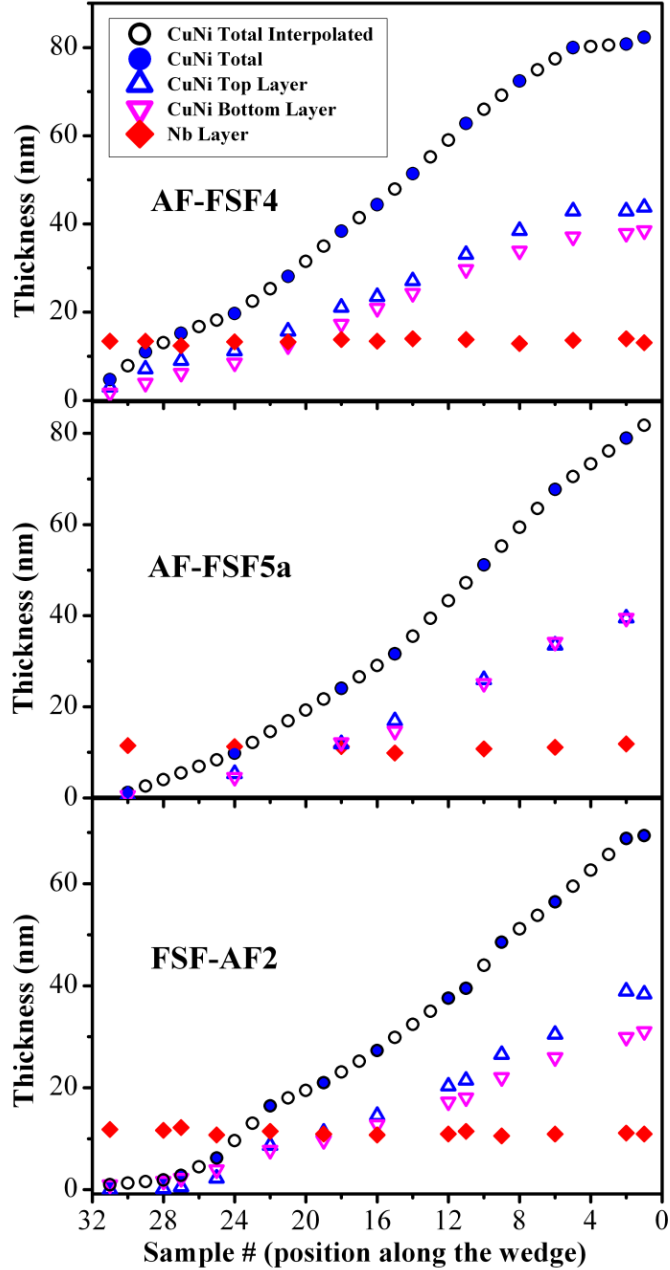

**Fig.3:** RBS measurements of the sample series. *Top and middle panels*: AF-FSF4 and AF-FSF5a series consisting of two ferromagnetic layers, which enclose a superconducting niobium layer of constant thickness, respectively, placed on a cobalt oxide layer of constant thickness of 7.7 and 8.4 nm in average (not shown here). *Bottom panel*: FSF-AF2 series consisting of two ferromagnetic layers, enclosing a superconducting niobium layer of constant thickness. On top of this stack there is an antiferromagnetic cobalt oxide layer with a thickness varying from 3.3 to 21 nm from sample #31 to #1 an average (not shown here).

In Fig. 3, bottom panel, the RBS results for sample series FSF-AF2 are shown, where the antiferromagnetic $CoO_x$ is now placed on the top of the F/S/F trilayer, obtained by a similar RBS fitting procedure as described above. Again, the two ferromagnetic layers adjacent to the superconducting Nb layer have a similar thickness profile. The average thickness of the Nb layer is 12.2 nm.

## C. Transmission Electron Microscopy

Transmission Electron Microscopy was done with a JEOL JEM-2100F microscope equipped with an Imaging Filter and a CCD camera from GATAN. In Fig. 4 the TEM image of sample #5 of series FSF-AF2 is shown. The individual layers can be clearly distinguished due to the sharp and straight interfaces between the different layers. The layers in this spin valve structure are, beginning from the substrate: $Cu_{49}Ni_{51}$/Nb/$Cu_{49}Ni_{51}$/$CoO_x$. The thickness of the layers stays constant over the range of the image. The boundary between the layers is sharply defined.

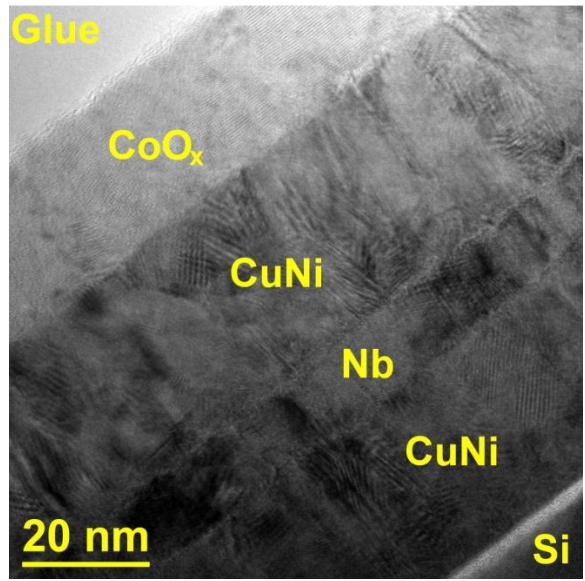

**Fig. 4:** Cross-sectional TEM picture of sample #5 of series FSF-AF2, consisting of a $Cu_{41}Ni_{59}$/Nb/$Cu_{41}Ni_{59}$ spin valve core structure with a cobalt oxide layer on top.

A similar TEM picture of sample #13 of series AF-FSF5a is shown in Fig. 5, where the $CoO_x$ layer is on the bottom of the F/S/F spin valve core structure, *i.e.* the sequence of the layers starting from the Si substrate is Si(buffer)/CoO$_x$/Cu$_{41}$Ni$_{59}$/Nb/Cu$_{41}$Ni$_{59}$/Si(cap). Again, the boundaries of the layers are sharp and smooth.

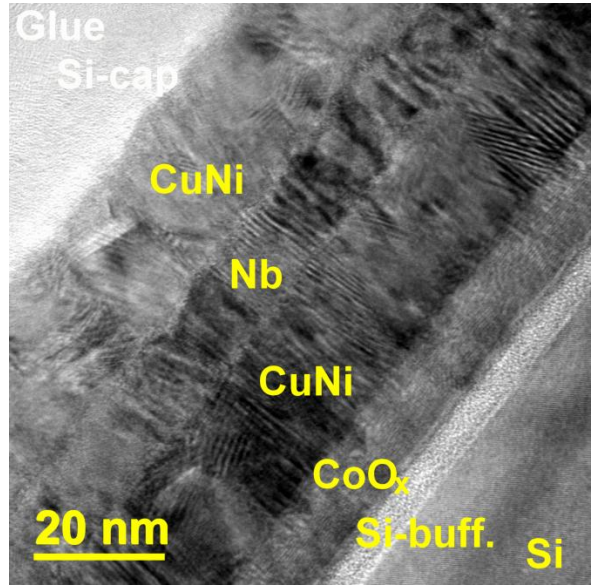

**Fig. 5:** Cross-sectional TEM picture of sample #13 of series AF-FSF5a, consisting of a Cu$_{41}$Ni$_{59}$/Nb/Cu$_{41}$Ni$_{59}$ spin valve core structure (protected by a Si-cap), on top of a cobalt oxide layer, deposited on a Si-buffer layer.

In order to get information about the lattice structure of cobalt oxide, and hence its oxidation state in sample series FSF-AF2, a high resolution image was taken within the cobalt oxide layer of sample #5 and the lattice spacings were measured. These spacings are 4.66 Å, measured between two adjacent lattice planes, and 2.83 Å, as can be seen in Fig. 6. During the deposition process it may have been formed CoO or Co$_3$O$_4$ (Co$_2$O$_3$ is not thermodynamically stable at ambient conditions [11]).

Cobalt monoxide has the rock salt NaCl structure, for which the lattice constant is 4.260 Å [12], whereas Co$_3$O$_4$ has a spinel structure with lattice constant 8.084 Å [13]. With these values the spacings of the {111} and {220} lattice planes can be calculated. In the case of CoO, the values for these planes are, 2.46 Å and 1.51 Å, whereas for Co$_3$O$_4$, one obtains 4.665 Å and 2.85 Å, respectively. By this analysis it can be concluded from the HRTEM image that, at least at this particular place of the sample, Co$_3$O$_4$ is present.

The lattice structure of cobalt oxide was checked over a larger area using electron diffraction on a different sample, consisting solely of cobalt oxide on top of the substrate, but sputtered with the same conditions. It results also in the evidence that $Co_3O_4$ is generated under the applied sputtering conditions [14].

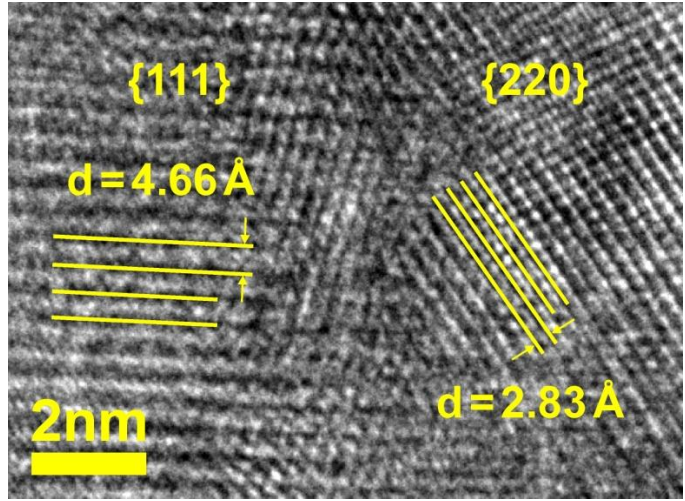

**Fig. 6.** HRTEM image of sample FSF-AF2 #5. The sample consists of a F/S/F spin valve core structure with F = $Cu_{41}Ni_{59}$, S = Nb, and with a $Co_3O_4$ layer on top. The image shows a section of the cobalt oxide. The indicated lattice planes belong to the {111} and {220} planes of $Co_3O_4$, respectively.

In Fig. 7 the cobalt oxide layer and its boundary to the bottom ferromagnetic layer of sample #13 of series AF-FSF5a are shown in high magnification. The cobalt oxide is deposited on the amorphous silicon buffer layer, covering the Si substrate. The lattice spacings of the substrate are visible, as well as some lattice planes of cobalt oxide. The lattice plane distances highlighted in the image in the cobalt oxide layer are 2.44, 2.46 and 2.08 Å, respectively. For exact determination of these values the scale was calibrated using the silicon high resolution region in the lower right corner of the image. The angles between these planes are 54° and 71°, respectively. From this, we conclude that the viewing direction is [110] with visible planes (002) and (111) of cubic CoO [15]. The plane distances for the CoO structure are d(002)=2.12 Å and d(111)=2.45 Å [11]. The value for the (111) distance fits well to the measured values, while the value for the (002) planes is by 1.9% too low, which may be due to strain within the layer or slight distortion of the image.

From this the presence of $Co_3O_4$ can not be ruled out. However, the characteristic plane distances of 5.7 Å or 4.6 Å typical for $Co_3O_4$ (found for other samples, see. e.g. Fig. 6) can not be found in this image or images taken at different regions on the same sample. Some of the measured plane distances can also be related to $Co_3O_4$, e.g. values about 2.13 Å or 2.43 Å [13]. Due to the fact that these planes are visible at regions without a low indexed viewing direction there is no additional information about angle relations between the visible planes and other planes, so it is not possible to conclude the cobalt oxide phase.

From this we can conclude that CoO is definitively present in the layer but we cannot rule out the presence of some $Co_3O_4$ completely.

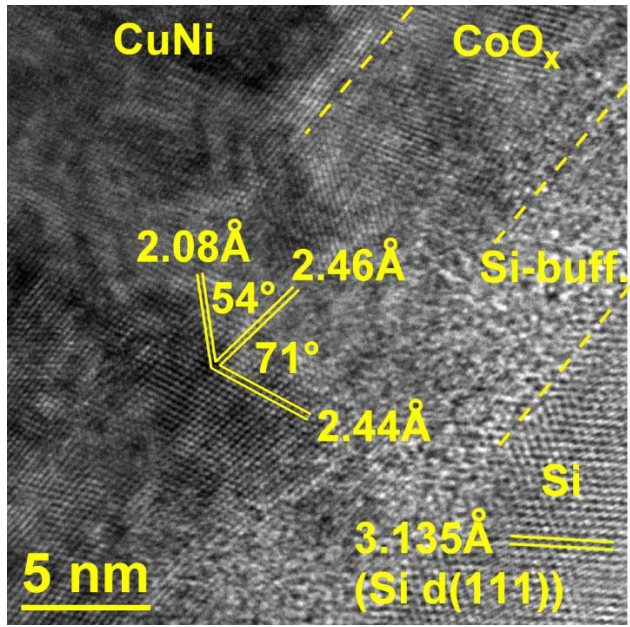

**Fig. 7:** Cross sectional HRTEM image of the lower part of sample #13 of series AF-FSF5a, consisting of the stack Si(buffer)/$CoO_x$/$Cu_{41}Ni_{59}$/Nb/$Cu_{41}Ni_{59}$/Si(cap).

**D. Magnetic Properties: Exchange Bias**

Another interesting property of the system is the exchange bias. It is a unidirectional anisotropy in the system which results from the interfacial exchange coupling of a ferromagnetic and an antiferromagnetic layer [10, 16]. The phenomenon is associated with a shift of the center of the magnetic hysteresis loop from the magnetic field $H=0$ to $H\neq0$ [17] as well as its broadening. The strength and direction of the exchange bias field $H_{EB}$ are determined by the cooling field $H_{FC}$ at which a sample is cooled down from above the Néel temperature of the antiferromagnet [18].

Antiferromagnetic alloys, such as FeMn and IrMn, often used to bias the ferromagnetic layers, induce a bias field $H_{EB}$ of the order of a few hundred Oersteds, what is not sufficient to achieve a complete antiparallel magnetic alignment of copper-nickel layers [19, 20]. Therefore we probed highly anisotropic cobalt oxide to exchange bias the $Cu_{41}Ni_{59}$ alloy layer.

To measure the hysteresis curve, a field of 1 T was applied at 300 K and the sample was cooled down. Hysteresis curves of sample #8 of the FSF-AF2 series measured with the field applied parallel and perpendicular to the film plane are shown in Fig. 8 as an example. So far, only a small exchange bias shift of about 20 Oe is present. The higher coercivity for the perpendicular direction indicates a dominating out of plane anisotropy in our $Cu_{41}Ni_{59}$ films.

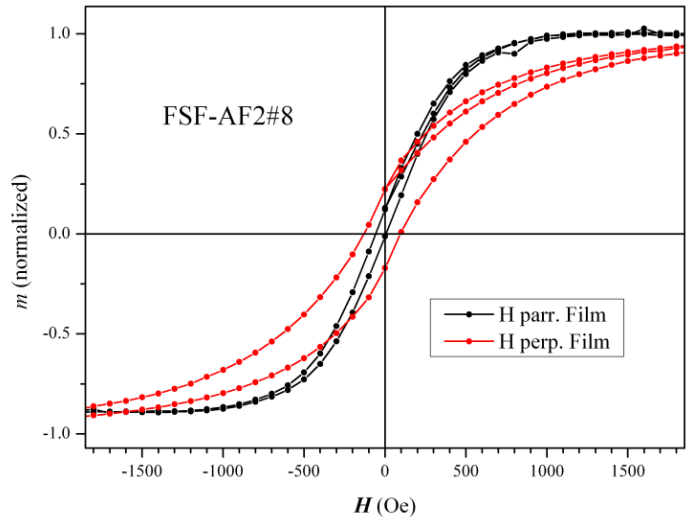

**Fig. 8:** Measurement of the hysteresis of sample #8 of the FSF-AF2 series. The normalized magnetic moment is plotted as a function of the applied magnetic field. The thicknesses of the copper-nickel alloy are 23 nm and 28 nm for the bottom and top layers, respectively

We attribute the small magnitude of the exchange bias to the magnetic easy axes mismatch between the $Cu_{41}Ni_{59}$ and the $CoO_x$ films [10, 21 – 23] (out-of-plane for $Cu_{41}Ni_{59}$ and in-plane for $CoO_x$). A route to exchange bias films with out-of-plane anisotropy could be, for example, to induce stress in $CoO_x$ to increase its anisotropy [24]. This is our next technological task.

## III. Results and Discussion
### A. Superconducting Properties

To determine the critical temperature of the samples, resistance versus temperature, $R(T)$, measurements have been performed in a $^3$He cryostat. Temperatures down to 1.4 K were realized by the 1-K stage, operated with liquid $^4$He. For lower temperatures (down to 380 mK) vapour pressure reduction of $^3$He was applied. The measuring current, in the standard four probe technique, was 10 µA and the polarity was changed during each measurement to avoid thermoelectric voltages. The samples were contacted by ultrasonic bonding with thin aluminium wires.

The superconducting transition temperature, $T_c$, is evaluated from the midpoint of the $R(T)$ curves. The transition width (0.1 $R_n$ to 0.9 $R_n$, where $R_n$, is the normal state resistance well above the transition) was below 0.2-0.3 K for most samples. The $T_c(d_F)$ measurements of sample series FSF-AF2 and AF-FSF4 have been plotted in Figs 9 and 10. They show a non-monotonic behaviour with a minimum at $d_{CuNi}^{Top} + d_{CuNi}^{Bottom}$ of about 15 nm. This is close to the double wedge geometry FSF spin valve core trilayer series FSF3 (with $d_{Nb}$=15.5 nm), for which the minimum of $T_c$ was observed for a thickness $d_{CuNi}^{Top} + d_{CuNi}^{Bottom}$ of 13.6 nm [3]. The minimum is deeper for the AF-FSF4 series compared to the FSF-AF2 and previous FSF3 series. Maybe, the smaller thickness of the Nb layer of the AF-FSF4 series (of $d_{Nb}$ = 13.4 nm) compared to the FSF3 series is the reason, although, the critical temperature at large $d_{CuNi}^{Top} + d_{CuNi}^{Bottom}$ is about 3 K in both AF-FSF4 and FSF3 series.

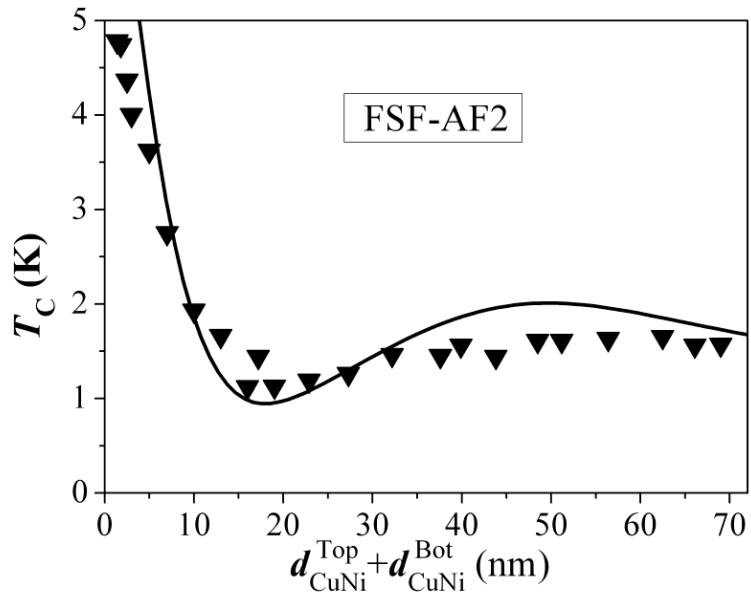

**Fig. 9:** Transition temperature measurement of the FSF-AF2 series as a function of the sum of the thicknesses of the bottom and top Cu$_{41}$Ni$_{59}$ alloy F-layers. The thickness of the niobium S-layer is 11.2 nm in average. The antiferromagnetic AF-layer is of CoO$_x$. The solid curve represents the results of modeling (see Section III.B).

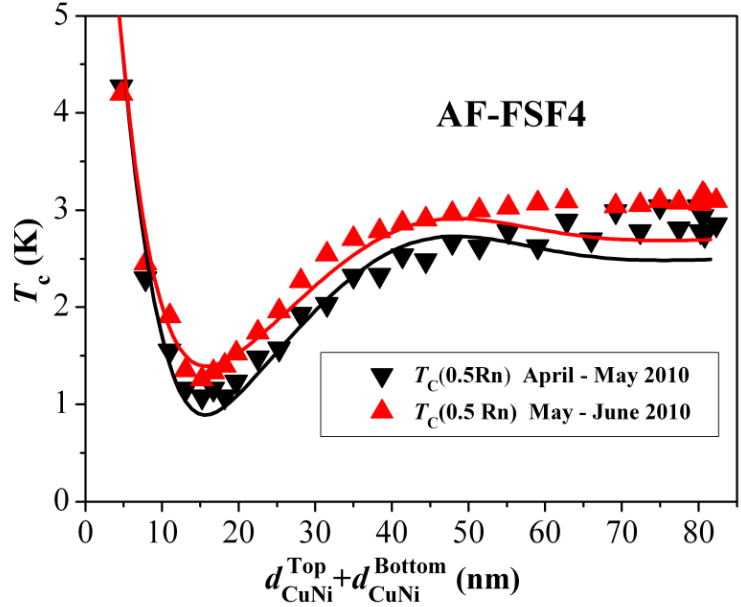

**Fig.10:** Transition temperature measurement of the AF-FSF4 series as a function of the sum of the thickness of the bottom and top $Cu_{41}Ni_{59}$ alloy F-layers. The thickness of the niobium S-layer is 13.4 nm in average. The antiferromagnetic AF-layer is of $CoO_x$. The solid curve represents the results of modeling (see Section III.B).

In the case of the FSF-AF2 series, the critical temperature at high copper-nickel thicknesses is only about 1.6 K. The reason may be that the thickness of the Nb layer is only about 11 nm in this range. However, a deeper minimum than observed is expected. The upper $Cu_{41}Ni_{59}$ and the $CoO_x$ layer on top were prepared while heating the specimen to about 200°C. Thus, an annealing of the sample occurred probably already during the deposition. Thus, aging effects similar to those discussed below, are the reason for the smooth minimum.

Since the F/S/F trilayer on top of the AF layer in sample series AF-FSF4 may be regarded as a mirror symmetric arrangement of a F/$\tilde{S}$ and $\tilde{S}$/F layer with S=2$\tilde{S}$, see Refs. [3, 9], one may introduce the average thickness $d_{\text{CuNi, AV}} = (1/2)(d_{CuNi}^{Top} + d_{CuNi}^{Bottom})$ and compare the results for $T_c$ with those for F/S and S/F bilayers with the half thickness of the Nb layer. In the first type of samples, the Nb layer is grown on top of the $Cu_{41}Ni_{59}$ film. It is vice versa for the S/F bilayers. For sample series AF-FSF4 and FSF-AF2, the minimum of $T_c$ is located at $d_{\text{CuNi, AV}} = 7.5$ nm, as typically observed for S/F bilayers [8].

There is a slight aging of the sample, increasing the temperature of the minimum of $T_c$ to slightly higher values. Since the position $d_{CuNi}^{Top} + d_{CuNi}^{Bottom}$ of the minimum does not change, this may indicate a slight decrease of the transparency of at least one of the interfaces between the superconductor and the ferromagnet, as will be discussed in the next section. Moreover, the limiting $T_c$ value at high copper nickel thicknesses is reached in a somewhat steeper manner.

In Fig. 11, the $T_c(d_F)$ measurements of sample series AF-FSF5a are shown, which were performed at three different times.

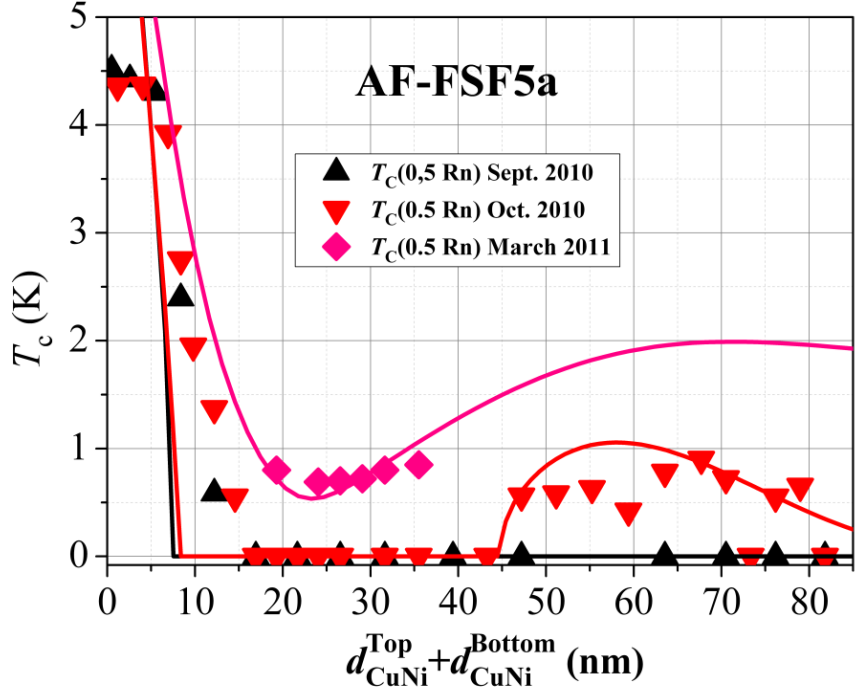

**Fig.11:**. Critical temperature measurement of the sample series AF-FSF5a as a function of the sum of the thicknesses of the bottom and top $Cu_{41}Ni_{59}$ alloy layers, measured at different times. The thickness of the niobium layer is 11.1 nm in average. The solid curve represents the results of modeling (see Section III.B).

The just prepared sample series (*Sep. 2010*) shows a steep decrease followed by an extinction of the superconducting transition temperature for $d_{CuNi}^{Top} + d_{CuNi}^{Bottom}$ above 16.9 nm. One month later (*Oct. 2010*) a reentrant superconducting behavior was observed. Finally (*March 2011*), the extinction region between 16.9 nm and 43.3 nm vanishes. The measured points indicate the change to an oscillating behavior with a minimum of the critical temperature at about 25 nm. For all measuring points drawn at zero critical temperature actually no superconducting transition has been observed for temperatures down to 380 mK, which is the lowest measuring temperature of our $^3$He cryostat. A $^3$He/$^4$He dilution refrigerator was used to investigate the samples from the extinction region at temperatures down to about 40 mK. However, at the meantime it appeared that the samples measured in March of 2011 have already changed to an oscillating critical temperature behavior with its minimum at about 0.7 K.

The observed reentrant behavior is similar to the double-wedge geometry series FSF5 investigated in Ref. [3]. The thickness of the niobium layer for that sample series was 12.8 nm, which means slightly thicker than for the AF-FSF5a series. The range of extinction region ($d_{CuNi}^{Top} + d_{CuNi}^{Bottom} \approx 17 - 48$ nm) is quite similar to the AF-FSF5a series, whereas the recovery of $T_c$, which was 1.3 K for the FSF5 series is somewhat higher. The minimum of the critical temperature oscillation in the *March 2011* measurements occurs at a higher copper nickel layer thickness, $(d_{CuNi}^{Top} + d_{CuNi}^{Bottom})_{minimum} \approx 23$ nm, as compared with the double wedge series FSF3 of Ref. [3] where it appears at 13.6 nm. Also the resulting value for the thickness per layer at $T_c$ minimum, $d_{CuNi, AV} = 23$ nm/2 = 11.5 nm, is significantly larger than observed for F/S and S/F bilayers [8, 9]. This observation will be further discussed at the end of Section. III.B, the aging effects observed are also discussed there in detail.

To probe the influence of magnetic configurations and local stray fields produced by the $Cu_{41}Ni_{59}$-alloy layers, we measured the magnetoresistance of our samples within the range of temperatures of the superconducting transition. A high magnetic field of 30 kOe was applied to the samples at a temperature of 300 K, above the Néel temperature of $CoO_x$ layer. Then the samples were cooled down to the measuring temperature to generate the exchange bias between the AF and the adjacent ferromagnetic layer. Then, the field was reduced to 6 kOe, and the magnetoresistance data were recorded at fixed temperature while the field was swept in the range of $\pm$ 6 kOe. The results of $R(H)$ measurements are presented in Fig. 12 for samples AF-FSF4#29 and FSF-AF2#23 for the field applied parallel to the thin films plane. The temperature was kept close to the midpoint of the superconducting transitions; particular values are given in the figure caption.

Signatures of a spin valve effect due to an antiparallel alignment of the magnetizations [8] or the generation of an odd-triplet component [23] were not observed. The reason that these effects are not present is probably a too weak exchange bias, so that the magnetization directions of both F-layers rotate simultaneously. Also in the case of an external magnetic field perpendicular to the surface of the layers of the sample no signature of the discussed spin valve effects are observed.

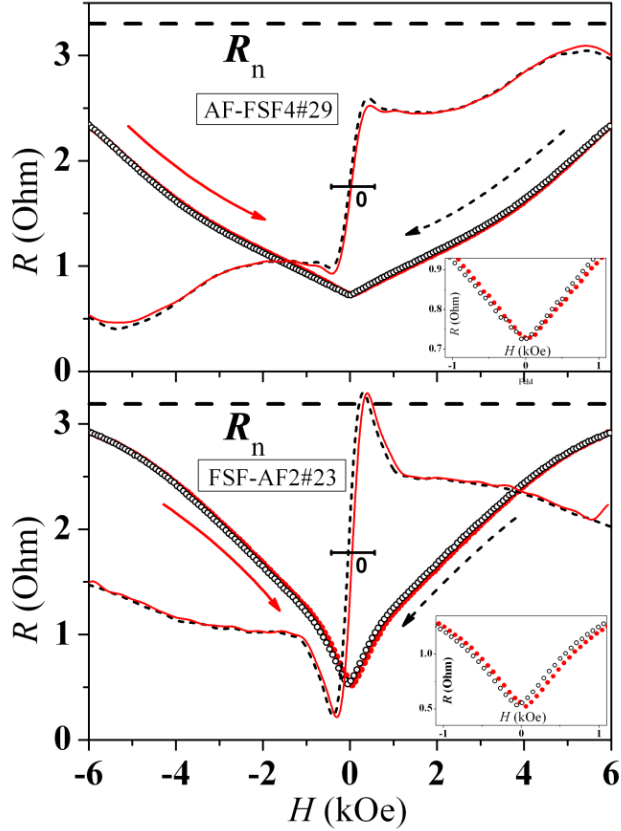

**Fig.12:** (a) $R(H)$ for sample AF-FSF4#29 at a temperature of 2 K; (b) $R(H)$ for sample FSF-AF2#23 at a temperature of 1.72 K. Here, $R_n$ indicated by the horizontal dashed line is the normal state resistance taken at 5 K. The direction of the magnetic field sweep is indicated by the arrows. The dashed black and solid red lines represent $dR/dH$ in arbitrary units.

The dip of the $R(H)$ curves around $H = 0$ Oe, extending over the typical width of the hysteresis curves between the coercive field and the saturation of our $Cu_{41}Ni_{59}$ thin films (Fig. 8), is probably generated by the effect of stray fields originating from the domain structure of the ferromagnetic layers. This effect usually leads to a reduction of the critical temperature of the superconducting layer [23, 25 – 29]. There is also the possibility of stray field enhanced superconductivity [30 – 32]. This effect is present here, indicated by a decrease of $R(H)$ in this range. The position of the minima of the $R(H)$ curves differs by about $\Delta H = 30$ Oe and 100 Oe for samples AF-FSF4#29 and FSF-AF2#23. This corresponds to the thickness of the $Cu_{41}Ni_{59}$ layers, *i.e.* this means the effect increases with the thickness. The order of magnitude of $\Delta H$ is comparable to the results given in Ref. [32], for $Nb/Cu_{43}Ni_{57}$ bilayers and $Cu_{43}Ni_{57}/Nb/Cu_{43}Ni_{57}$ trilayers.

## B. Discussion of the superconducting properties

According to the theory of Bardeen-Cooper and-Schrieffer (BCS) [33, 34], the superconducting state is buildup of pairs of electrons (Cooper pairs) with opposite spin and wave number vectors, *i.e.* momenta of the electrons. Since ferromagnetism requires a parallel orientation of the spins, both ordered states are antagonistic. Nevertheless, Fulde-Ferrell [3] and Larkin-Ovchinnikov [4] predicted the existence of a superconducting state in a ferromagnetic material, *i.e.* in the presence of an exchange field. In this state a singlet s-wave pairing is present with opposite spin and oppositely directed wave number vectors of the electrons, as in the BCS case. However, the absolute values of the wave number vectors are different resulting in non-zero pairing momentum (see extensive discussion in [6 – 9, 35]).

The behavior of the transition temperature of a F/S/F spin valve core structure is the result of interference effects of the oscillating FFLO like state pairing function, leading to oscillations of $T_c$ or even reentrant behavior of superconductivity, where there is an extinction and recovery of the superconducting state, as a function of the F-layer thickness. Both phenomena have been observed in S/F and F/S bilayers [7 – 9, 36] and in a F/S/F trilayers [3] utilizing copper-nickel alloy as F-layer material.

If now one of the F-layers is pinned by an AF layer, an applied magnetic field may be used to turn around the magnetization directions of the other F-layer, thus generating the triplet zero projection and odd in frequency triplet state, and thereby changing the transition temperature [37, 38]. That means that one has a superconducting spin valve. As we discussed in detail in Ref. [8], the spin-valve effect is expected to be most expressed for trilayers with reentrant superconducting behavior, however, the exchange bias of about 20 Oe which we obtained in our samples is too small to produce sizeable spin-valve effect.

The theoretical curve fitted to the experiments of the present work are calculated by extending the dirty case theory to the clean case of a ferromagnet ($l_F \gg \xi_{F0}$, the superconductor is always assumed to be in the dirty limit) and applying it to our samples which are between the clean and the dirty limit, see Refs. [3, 8, 39] and Section 2.2 of [14]. To fit the theory we put the magnetization directions of the magnetic layers of our AF-FSF and FSF-AF systems parallel considering the lack of exchange bias in our magnetic measurements (Sec. II.D, see also more detailed discussion at the end of this subsection). The experiments are well described by the theory (which in this case only considers the FFLO singlet state [37, 38]) using the following parameters.

**Table 1**: Overview of the fitting parameters used for the curves of the theoretical predictions of the transition temperature of the sample series AF-FSF4, AF-FSF5a, and FSF-AF2

| Sample series | $T_{c0,Nb}$ ($d_{CuNi}=0$ nm) | $\xi_S$ | $N_F v_F/N_S v_S$ Top/Bottom | $T_F$ Top/Bottom | $l_F/\xi_{F0}$ Top/Bottom | $\xi_{F0}$ Top/Bottom |
|---|---|---|---|---|---|---|
| AF-FSF 4 | 7.8 K | 6.2 nm | 0.20/0.18 | 0.63/0.78 | 0.8/0.62 | 10.4/10.2 nm |
| Aged | 7.8 | 6.2 nm | 0.20/0.18 | 0.63/0.71 | 0.8/0.62 | 10.4/10.2 nm |
| AF-FSF5a | 7.4 K | 6.2 nm | 0.20/0.18 | 0.66/0.78 | 0.90/0.65 | 10.4/10.2 nm |
| Aged1 | 7.4 K | 6.2 nm | 0.20/0.18 | 0.63/0.70 | 0.90/0.65 | 10.8/10.2 nm |
| Aged2 | 7.0 K | 6.2 nm | 0.20/0.18 | 0.62/0.51 | 0.5/0.4 | 17.0/16.0 nm |
| FSF-AF2 | 6.8 K | 6.2 nm | 0.195/0.18 | 0.54/0.66 | 0.60/0.50 | 12.2/11.8 nm |

.

The physical parameters are explained in detail in Ref. [8]: $\xi_S$ the superconducting coherence length, $\xi_{F0}$ the coherence length for Cooper pairs in a ferromagnetic metal, $l_F$ the mean free path of conduction electrons in the ferromagnetic material, $N_F v_F/N_S v_S$ the ratio of Sharvin conductances at the S/F interface, and $T_F$ the interface transparency parameter. The critical temperature $T_{c0}$ of the stand alone niobium layer is obtained from Fig. 5 of Ref. [8]. In the case of the aged samples of the AF-FSF5a series and the annealed FSF-AF2 series a further reduction of the critical temperature is assumed.

For samples series AF-FSF4 and AF-FSF5a, where aging effects are observed, a change of the transparency (generated probably by oxygen diffusing from the $CoO_x$ to the bottom $Cu_{41}Ni_{59}$/Nb boundary, yielding an insulating $NbO_x$ layer [40, 41]) is assumed. Moreover, in curve *March 2011* from sample series AF-FSF5a, the minimum of $T_c$ seems to appear at an unusually high value of about 23 nm for $d_{CuNi}^{Top} + d_{CuNi}^{Bottom}$ (compared to previous measurements on F/S/F trilayers [3]). If we divide by 2, we get 11.5 nm which is higher than the values observed for S/F and F/S bilayers of 7.5 nm and 10 nm, respectively. As we discussed in detail in Ref. [9], this indicates an increase of $\xi_{F0}$, i.e. a decrease of $E_{ex}$, which may be caused by a degradation of the $Cu_{41}Ni_{59}$ alloy (possibly by converting Ni to $NiO_x$, which is antiferromagnetic [42]).

To reduce the aging problems, a thin niobium or aluminum interlayer (thickness of the order of 1 nm) between the $CoO_x$ and F-material may be introduced. Such layer is expected to capture the diffusing oxygen by being converted to niobium oxide or aluminum oxide due to a solid state reaction. Such oxide layer at the same time may serve to increase the exchange bias (see the discussion in [43]).

## IV. Conclusion

In the present work we demonstrated that F/S/F spin valve core structures (F=$Cu_{41}Ni_{59}$, S=Nb) can be experimentally realized on an antiferromagnetic $CoO_x$ layer. In a spin-valve this layer serves to exchange bias one of the F-layers against the rotation of the magnetization direction in an external magnetic field. The magnetic measurements exhibited exchange bias in our system, however, probably due to its insufficient value (about 20 Oe), signatures of spin-valve effects have not been detected.

Detailed investigations of the deposited thin films by Rutherford Backscattering Spectrometry yielded a precise knowledge of the layer thicknesses. High resolution Transmission Electron Microscopy was applied to demonstrate the high quality of the interfaces between the layers and to investigate the oxidation state of $CoO_x$ layer.

The core structures showed deep critical temperature oscillations and reentrant superconducting behavior, as required for a considerable spin switching effect. The superconducting properties of the spin-valve core could be successfully fitted by the theory. This demonstrates that the FFLO-like state in our F/S/F core structure, grown on $CoO_x$ antiferromagnetic material, retained undisturbed.

**Acknowledgments**

The authors are grateful to S. Heidemeyer, B. Knoblich and W. Reiber for assistance in the TEM sample preparation. The work was supported by the Deutsche Forschungsgemeinschaft (DFG) under the grant No GZ: HO 955/6-1,2. In part the Russian Fund for Basic Research (RFBR) supported the project under the grant 11-02-00848-a (LRT).